\documentclass[a4paper,10pt]{extarticle}
\usepackage{geometry}
\usepackage{breqn}
\usepackage[usenames,dvipsnames]{color}
\usepackage{indentfirst}
\usepackage[utf8]{inputenc}
\usepackage[T1]{fontenc}
\usepackage{amsmath}
\usepackage{mathtools}
\usepackage[thinc]{esdiff}
\usepackage[bottom]{footmisc}
\usepackage{float}
\usepackage[euler]{textgreek}
\usepackage{siunitx}
\usepackage{subfig}
\usepackage{graphicx}
\usepackage{dcolumn}
\usepackage{bm}
\usepackage{braket}
\usepackage{physics}
\usepackage{xfrac}
\usepackage{authblk}
\usepackage{blindtext}

\begin{document}
\title{A Toy Model for Low Energy Nuclear Fusion}
\date{June 2022}
\author[1]{K. Ramkumar\thanks{ramkumar@iitk.ac.in}}
\author[1]{Harishyam Kumar\thanks{hari@iitk.ac.in}}
\author[2]{Pankaj Jain\thanks{pkjain@iitk.ac.in}}
\affil[1]{Department of Physics, Indian Institute of Technology, Kanpur}
\affil[2]{Department of Space Science \& Astronomy, Indian Institute of Technology, Kanpur}

\maketitle

\begin{abstract}
We study the fusion of a proton with a nucleus with the emission 
of two photons at low incident energy of the order of eV or smaller.
We use a step model for the repulsive potential between proton and the 
nuclei. We consider the reaction both in free space and inside
a medium. We make a simple model for the medium 
	by assuming a hard wall potential beyond
a certain length scale. This essentially leads to discretization of
the energy spectrum which is expected inside a medium and is seen both for 
a crystalline lattice structure and for amorphous materials. 
We use second order perturbation theory to compute the transistion
rate. 
	We find that
the rate in free space is very small. However in medium, the rate may
be substantial. Hence, we conclude that nuclear fusion reactions 
	may take place at low energies at observable rates. 
\end{abstract}

\section{Introduction}
The nuclear fusion processes are expected to be strongly suppressed
at low energies \cite{1968psen.book.....C}. There have been experimental 
claims that such processes
may be occuring at observable rates in a medium, see for example \cite{doi:10.1002/9781118043493.ch41,doi:10.1002/9781118043493.ch43,StormsCS2015,Biberian19}. However, despite considerable effort \cite{SinhaCS2015,Celani17,SPITALERI2016275,Hagelstein19,Meulenberg19}, so far there
does not exist any reliable theoretical model of how this can happen.
A useful review of the shortcomings of a wide range of theoretical proposals 
is given in \cite{Chechin_1994}.

In \cite{Jain2020,Jain2021}, the authors have explored the possibility that low energy
fusion may arise at second order in perturbation theory. 
A similar idea has also been proposed in \cite{PhysRevC.99.054620}. At this order we 
need to sum over all the intermediate state energies and hence the 
Coulomb barrier may not be very prohibitive. The 
process considered in \cite{Jain2020,Jain2021} 
involves two interactions or vertices and involves emission of two
photons, one at each vertex. The process may be expressed as,
\begin{equation}
	{}^1H + {}^A{\rm X} \rightarrow {}^{A+1}{\rm Y} +\gamma(\omega_1) + \gamma(\omega_2)
	\label{eq:twophotons}
\end{equation}
where ${}^1H$ denotes the Hydrogen, ${}^A{\rm X}$ a nucleus with atomic number $Z$ and
mass number $A$ and ${}^{A+1}Y$ a nucleus with atomic number $Z+1$ and mass
number $A+1$. The emitted photons have frequencies $\omega_1$ and $\omega_2$.
This may be compared to the related first order process,
\begin{equation}
	{}^1H + {}^A {\rm X} \rightarrow {}^{A+1}{\rm Y} +\gamma(\omega)
	\label{eq:standard}
\end{equation}
which is the standard fusion process with emission of a photon.

\begin{figure}[H]
\centering
\includegraphics[ clip,scale=0.5]{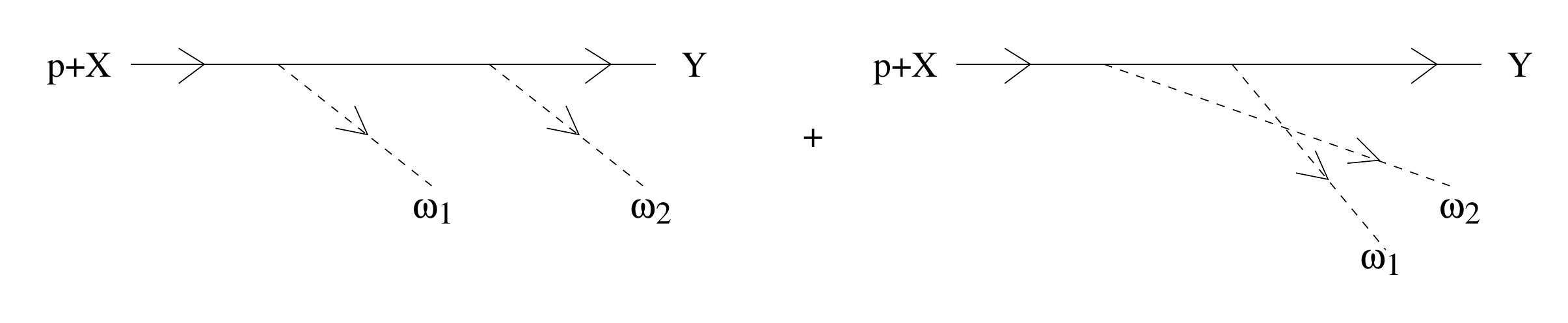}
\caption{\label{fig:twophotons} The two amplitudes contributing to the
	nuclear fusion process with two photon emission.  
	}
\end{figure}

At second order in perturbation theory, the two photon emission process, Eq. \ref{eq:twophotons} gets contributions from two amplitudes which are shown 
in Fig. \ref{fig:twophotons}. At the first interaction or vertex the proton
or the X nucleus
emits a photon forming an intermediate state consisting of a proton and X.
We work in the center of mass and relative coordinates and only the 
relative coordinates are relevant \cite{Jain2020,Jain2021}.
In the intermediate state we need to sum over states of all energy, without
imposing energy conservation at either of the two vertices. Of course, 
the total energy has to be conserved.
At the second vertex, the proton
gets captured by the nucleus X with emission of another photon. 
The 
process, therefore, involves two matrix elements one for each vertex. In
earlier papers \cite{Jain2020,Jain2021}, the authors
refer to the first matrix element
as the molecular matrix element since it gets dominant contributions from
distances of order 1 atomic unit, while
the second matrix elements gets contributions dominantly from nuclear 
distances and is called nuclear matrix element. 

The important point is that we need to sum over intermediate states of
all energies. 
Since the initial
state has very small energy $E_i$ and momentum, 
the molecular matrix element is appreciable
only when the energy of the intermediate state $E_n$ is such that the corresponding
momentum $\vec P_n$ closely balances the photon momentum $\vec P_\gamma$. 
One might expect that the dominant contribution to the entire amplitude
would come from $\vec P_n\approx P_\gamma$ with the corresponding energy 
$E_n>>E_i$, thereby leading to a rather large amplitude. However, explicit
calculations in \cite{Jain2020,Jain2021} show that this fails. The problem
is that here we are dealing with particles in a potential and hence the
energy eigenstates are not eigenstates of momentum. Due to this the molecular
matrix element does not select a unique value of $\vec P_n$ and a large
range of values of $P_n = |\vec P_n|$ contribute. Explicit calculations show
that these cancel among one another leading to a very small amplitude. 

In \cite{Jain2020,Jain2021}, the authors suggested several possibities which might
evade the acute cancellation described above. One 
suggestion was that in a medium, the energy eigenvalues are discrete and do
not form a continuous set as in free space. In such a case, the cancellation
may not be complete and we may get a substantial contribution. We test
this possibility in detail in the present paper, 
assuming a step function potential for
the tunneling barrier. This captures the essential details of the process
while facilitating mathematical calculations. Calculation with Coulomb
is postponed to future work. 
Another suggestion in
\cite{Jain2020,Jain2021} was the
 presence of a resonant nuclear state at energy $E_R>> E_i$.
In this case, the dominant contribution will arise from energies very
close to $E_R$ which may not cancel out. 
We briefly comment on this possibility also in the present paper.

We also point out that in our paper we have confined ourselves to one
process in order to demonstrate that low energy nuclear reactions are 
possible. There may be other processes which may not involve emission
of photons and might proceed at higher rates. This needs to be studied
in detail in future. While demonstrating the theoretical possibility
of such reactions to occur at observable rates, our work paves the way 
for study of other related processes. 

\section{Potential Model and Wave functions}
\label{sec:model}

In this section, we give the potential model used for the calculations
and the corresponding energy eigenstates. As mentioned in the Introduction
we use a step function for the potential barrier instead of the
standard Coulomb potential. This has the advantage that the wave fucntions 
can be computed analytically.

\begin{figure}[H]
\centering
\includegraphics[ clip,scale=0.7]{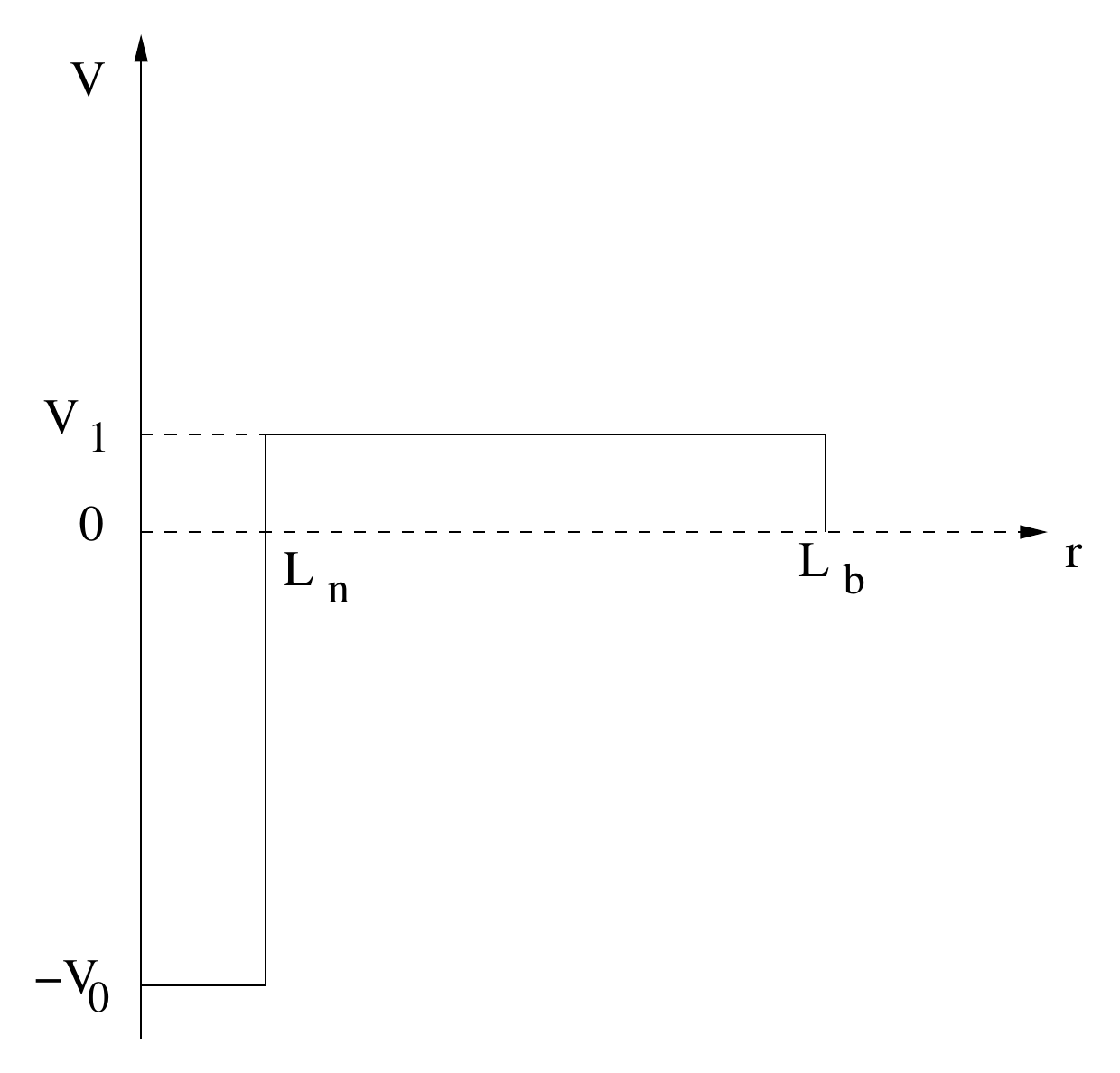}
\caption{\label{fig:pot} Schematic illustration of the model potential.
}
\end{figure}

We assume a spherically symmetric potential which can be written as, 
\begin{equation}
V(r) = \begin{cases} 
      -V_0 & r < L_n \\
      V_1 & L_n\leq r\leq L_{b} \\
      0 & L_c\ge r > L_{b} \\
      V_2 & r > L_c
   \end{cases}
	\label{eq:potential}
\end{equation}
where $V_0$ is the nuclear potential, $V_1$ represents our model 
for the tunneling barrier and we shall take $V_2$ to be infinitely large. The potential is shown schematically in Fig.
\ref{fig:pot}. 
A possible set of parameters are $V_0=50 $ MeV, $V_1=100$ in atomic
units (Hartree), the nuclear length scale
$L_n=\num{0.9566e-4}$ atomic units (a.u.) 
and the barrier length scale $L_b =
0.1$ a.u.. We point out that 1 Hartree is approximately 27.2 eV
and the atomic unit for length is Bohr radius. The cutoff length scale $L_c$ is not shown in Fig. \ref{fig:pot}. We assume that the potential $V_2$ beyond this point is infinitely large. Hence all the wave functions are set to zero at this radius.

The potential model represents a mathematically well defined quantum system. It also approximately describes a physical situation in which we have a small spherically symmetric cavity in a solid material composed of element with large atomic mass. In the cavity we assume presence of a gas composed of Hydrogen and a relatively heavier element X. We are considering fusion of Hydrogen with X. The effect of solid matter is to simply discretize the energy eigenvalues of the H$-$X system.  Each lattice site of the solid matter presents a very high potential barrier to H$-$X system.  We assume that this barrier height $V_2>>V_1$ and,
for simplicity, we approximate it as infinity. If this forms a regular lattice, we expect the formation of bands as in the case of an electronic system
\cite{Ashcroft76}. Alternatively if there is sufficient disorder in the system, we may only have a few localized states, analogous to Anderson localization \cite{doi:10.1142/7663,RevModPhys.57.287}. Here we have simply modelled this system by assuming that the overall effect of the lattice sites is to present an almost infinite potential barrier once we go to sufficiently large radius from the cavity. This will lead to discretization of the energy levels which is qualitatively similar to the band structure we expect for a crystalline lattice,
 in the sense that both lead to non-continuous quantum states. 

The potential given in Eq. \ref{eq:potential} can be solved analytically. 
We are interested in the two particle wave function. The center of mass
wave function is not relevant to our analysis \cite{Jain2020,Jain2021} and
we focus on the relative coordinate $\vec r$. For $l=0$ the wave function
depends only on $r=|\vec r|$ and can be expressed as,
\begin{equation}
	\psi(r)=Y_0^0 {U(r)\over r} \,.
\end{equation}
where $Y_0^0=1/\sqrt{4\pi}$.
For energy eigenvalue $E<V_1$, we obtain,	
\begin{subequations}\label{eq:wavefunc}
\begin{gather}
	U(r) = \frac{1}{N'_L}\ \sin\ (K_1\ r) \ \ \ \ \ \ \ \ \ \ \ \ \ \ \ \  {\rm for}\ r<L_n\\
	U(r) = \frac{1}{N'_L}\ \left[ B_L\ e^{-K_2 r} + C_L\ e^{K_2 r}\right]\ \ \ \ \ \ \ \ \ \ {\rm for}\ L_n\leq r\leq L_{b}\\
	U(r) = \frac{1}{N'_L}\ \left[ D_L\ \sin(K\ r) + F_L\ \cos(K\ r)\right]
	\ \ \  {\rm for}\ L_b< r\le L_{c} \\
	U(r) = 0
	\ \ \ \ \ \ \ \ \ \ \ \ \ \ \ \ \ \ \ \ \ \ \ \ \ \ \ \ \ \ \ \ \ \ \   {\rm for}\ r>L_{c} 
    \end{gather}
\end{subequations}
where,
\begin{equation}
    K = \sqrt{\frac{2mE}{\hbar^2}}, \ \ \
    K_1 = \sqrt{\frac{2m(E-V_0)}{\hbar^2}}, \ \ \
    K_2 = \sqrt{\frac{2m(V_1-E)}{\hbar^2}},
	\label{eq:knumbers}
\end{equation}
\begin{equation}
    N'_L=K\ N_L,  \ \ \ 
    N_L=\sqrt{D_L^2+F_L^2},
	\label{eq:norm}
\end{equation}
\begin{equation}
    B_L = \frac{b}{2\ K_2\ e^{-K_2\ L_n}},\ \ \ 
    C_L = \frac{c }{2\ K_2\ e^{K_2\ L_n}},
	\label{eq:coeffBC}
\end{equation}
\begin{eqnarray}
	b &=& -K_1\ \cos(K_1\ L_n) + K_2\ \sin(K_1\ L_n)\\
	c &=& {K_1\ \cos(K_1\ L_n) + K_2\ \sin(K_1\ L_n)}
	\label{eq:coeffbc}
\end{eqnarray}
and
\begin{subequations}\label{eq:h2}
\begin{gather}
    D_L= (B_L\ e^{-K_2 L_b} + C\ e^{K_2 L_b})\ \sin(K\ L_b) - \frac{K_2}{K}\ (B_L\ e^{-K_2 L_b} - C_L\ e^{K_2 L_b})\ \cos(K\ L_b)\label{eq:dd}\\
    F_L= (B_L\ e^{-K_2 L_b} + C_L\ e^{K_2 L_b})\ \cos(K\ L_b) + \frac{K_2}{K}\ (B_L\ e^{-K_2 L_b} - C_L\ e^{K_2 L_b})\ \sin(K\ L_b) \label{eq:ee}
\end{gather}
\end{subequations}
Furthermore we have
\begin{equation}
	D_L\sin(KL_c) + F_L\cos(KL_c) = 0
\end{equation}
Due to the boundary condition at $L_c$ we get discrete energy eigenvalues
and each eigenfunction is normalized to unity.

For $E\ge V_1$, the eigenfunctions for $r<L_n$ and $r>L_b$ 
have the same form as for $E<V_1$ with $N'_L$, $D_L$ and $F_L$ replaced by
$N'_H$, $D_H$ and $F_H$ respectively. In the region
$L_n\le r\le L_b$, it takes the form,
\begin{equation}
U(r) = \frac{1}{N'_L}\ \left[ B_H\ \sin(\tilde K_2 r) + C_H\ \cos(\tilde K_2 r)\right]\ \ \ \ \ \ \ \ \ \ {\rm for}\ L_n\leq r\leq L_{b}\\
\end{equation}
with 
\begin{equation}
	\tilde K_2 = \sqrt{2m(E-V_1)\over \hbar^2}
\end{equation}
The coefficients are given by
\begin{eqnarray}
	B_H & =& {K_1\over \tilde K_2} \cos(K_1L_n) \cos(\tilde K_2L_n)
	+\sin(K_1L_n) \sin(\tilde K_2L_n)\nonumber\\
	C_H & =& -{K_1\over \tilde K_2} \cos(K_1L_n) \sin(\tilde K_2L_n)
	+\sin(K_1L_n) \cos(\tilde K_2L_n)\nonumber\\
	D_H &=& \left(B_H\sin(\tilde K_2 L_b ) + C_H\cos(\tilde K_2 L_b)\right) \sin(KL_b)\nonumber\\
	&+& {\tilde K_2\over K} \left(B_H\cos(\tilde K_2 L_b ) - C_H\sin(\tilde K_2 L_b)\right)\cos(KL_b)\nonumber\\
	F_H &=& \left(B_H\sin(\tilde K_2 L_b ) + C_H\cos(\tilde K_2 L_b)\right) \cos(KL_b)\nonumber\\
	&-& {\tilde K_2\over K} \left(B_H\cos(\tilde K_2 L_b ) - C_H\sin(\tilde K_2 L_b)\right)\sin(KL_b)
\end{eqnarray}


The final state wave function is taken to be $l=1$. 
In our calculation we take a particular initial and final state. This is 
sufficient to determine whether fusion is facilitated by the second order
mechanism. 
The detailed angular
and spin dependence of the wave functions is given below. 

In our analysis we shall assume that the initial state energy $E_i$ is very
small, of order 0.1 eV. For such energy  
the standard leading order process Eq. \ref{eq:standard} 
is very strongly suppressed since the initial state wave function is
neglible at small nuclear distances. 
 The second order process Eq. \ref{eq:twophotons}, however,
gets contributions from all energy eigenvalues in the intermediate state.
For large energy, the corresponding eigenstates may take large values 
at small distances. 
 Hence it is possible that the corresponding
amplitude may be large.

\section{Reaction Rate at First Order }

\begin{figure}[H]
\centering
\includegraphics[ clip,scale=0.6]{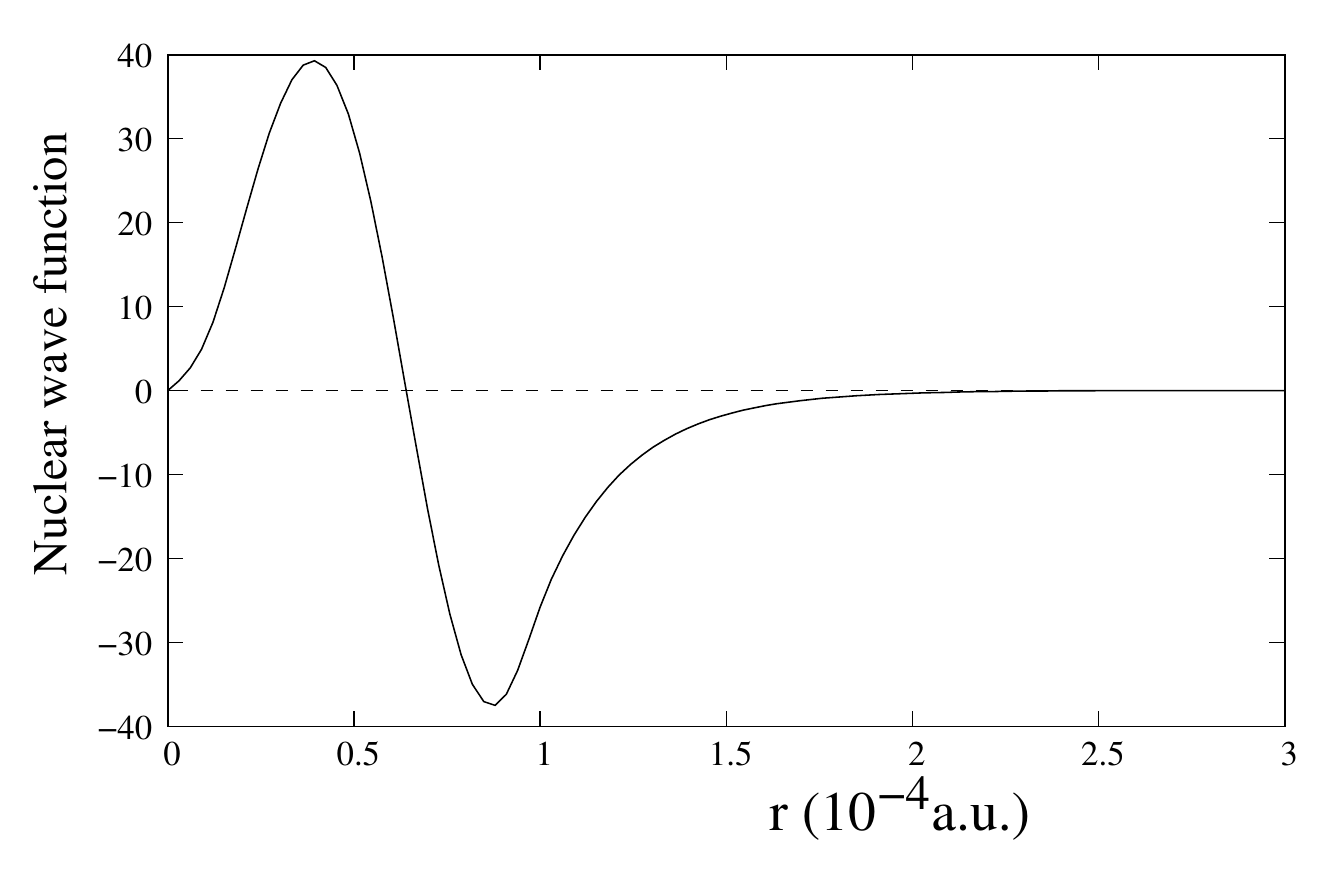}
\caption{\label{fig:nuc} The final state nuclear wave function ($l=1$) 
	corresponding to energy eigenvalue -13.5 MeV.}
\end{figure}

In this section we compute the cross section for the leading order
process given in Eq. \ref{eq:standard}. This is useful for comparison with
the second order resonant process, discussed in the next section. 
We assume that the nucleus ${}^A{\rm X}$ has spin 0 and $l=0$ in the ground
state. We consider the transition from initial state composed of 
 ${}^1H$ and ${}^A{\rm X}$
with $l=0$ to the final state ${}^{A+1}{\rm Y}$
with $l=1$. The spin wave function of proton does not play any role 
in this transition. 
For the chosen potential parameters, the final state has energy
eigenvalue $E_f=-13.5$ MeV and the corresponding wave function is shown
in Fig. \ref{fig:nuc}.

We may write the Hamiltonian as,
\begin{equation}
    H  = H_0 + H_I
\end{equation}
where $H_0$ is the unperturbed Hamiltonian and $H_I$ the electromagnetic
perturbation. We can write $H_0$ as,
\begin{equation}
	H_0 = {\cal K}_1 + {\cal K}_2 + { V}(r)
\end{equation}
where ${\cal K}_1$ and ${\cal K}_2$ are the kinetic energies of the ${}^1H$ and the ${}^AX$
nucleus respectively and ${ V}(r)$ the potential given in Eq. 
\ref{eq:potential}. 
Here we focus on the relative motion relevant for
the fusion process. The relative coordinate is denoted by $\vec r$, 
such that, $\vec r=\vec r_2-\vec r_1$, where $\vec r_1$ and $\vec r_2$ 
are the coordinates of the ${}^1H$ and ${}^A{\rm X}$ nucleus. 
The potential is assumed to
be spherically symmetric and depends only on the magnitude $r=|\vec r|$. 
We express the kinetic energies in terms of the center of mass and 
relative momenta and focus on the relative momentum. 

The perturbation 
 $H_I$ can be expressed as \cite{merzbacher1998quantum,sakurai1967advanced}
 \begin{equation}
    H_I(t) = -{ Z_1e\over cm_1} \vec A(\vec r_1,t)\cdot \vec p_1 - { Z_2e\over cm_2} \vec A(\vec r_2,t)\cdot \vec p_2 + {e\hbar g_p\over 2 m_1c} \vec\sigma\cdot \vec B + ...
    \label{eq:Hint}
\end{equation}
where
$Z_1$, $m_1$ and $\vec p_1$  are the atomic number, mass and momentum of 
the ${}^1H$. The corresponding quantities for ${}^A{\rm X}$ are
$Z_2$, $m_2$ and $\vec p_2$. Furthermore, 
$\sigma_i$ are the Pauli matrices, $g_p$ the proton $g$
factor, 
 $\vec A$ is the vector potential, 
\begin{equation}
	\vec A(\vec r,t) = {1\over \sqrt{\Omega}} \sum_{\vec k}\sum_\beta c\sqrt{\hbar\over 2\omega} \left[a_{\vec k,\beta}(t) \vec\epsilon_\beta e^{i\vec k\cdot\vec r} +  a^\dagger_{\vec k,\beta}(t) \vec\epsilon_\beta^{\,*} e^{-i\vec k\cdot\vec r}\right]
\end{equation}
and
$\vec B=\vec \nabla\times \vec A$
is the magnetic field. The vector potential is expressed in terms of the
photon polarization vector $\epsilon_\beta$, the wave vector $\vec k$ 
the frequency $\omega$ and the total volume $\Omega$.  

We next compute the rate for $l=0$ to $l=1$ transition at first
order in perturbation theory. We consider a particular process in which 
the initial state proton is in spin up state. The final state is taken to be
$j=3/2$ and $j_z=3/2$. The corresponding rate is given by, 
\begin{equation}
\Gamma_1 = \frac{4\alpha\xi^2 E_p^3}{3\hbar^3c^2}|\bra{\psi_f}r\ket{\psi_i}|^2
\end{equation}
where $E_p$ is the energy of the photon emitted as given by the conservation of energy,
\[E_p=E_f-E_i\]
We evaluate the reaction rate 
by setting the cutoff length scale $L_c=10$ atomic units. For such a length
scale we find an eigenstate 
with energy equal to  $E_i=0.0903$ eV. 
For the model potential we are considering, 
the reaction rate for this state is found to be $10^{-44}$ per second.

\subsection{Transition Rate at Second Order}
In this section we compute the transition rate for the process given in Eq. 
\ref{eq:twophotons} which goes by emission of two photons. This gets dominant
contribution at second order in perturbation theory.
The transition amplitude in this order is given by,
\begin{eqnarray} \label{eq:transition_amp}
	\bra{f}T(t_0,t)\ket{i} &=& \left(\frac{-i}{\hbar}\right)^2 \sum_n \int_{t_0}^{t} dt'\: \bra{f}e^{i\:H_0t'/\hbar}H_{I}(t')e^{-i\:H_0t'/\hbar}\ket{n}\nonumber\\ 
	&\times &\int_{t_0}^{t'} dt''\: \bra{n}e^{i\:H_0t''/\hbar}H_{I}(t'')e^{-i\:H_0t''/\hbar}\ket{i}
\end{eqnarray}
where $H_I$ is given by Eq. \ref{eq:Hint}. 
As mentioned above, 
we take the initial state to be $l=0$, spin $1/2$ with $S_z = -1/2$,
corresponding to the spin of the proton.  
For the first transition from initial to intermediate state we consider contribution from the 
magnetic term in Eq. \ref{eq:Hint}. This transition does not change $l$ but
flips the spin of the proton. The transition from intermediate to final
state is taken to be $l=0$ to $l=1$ with no change in spin. This gets dominant
contribution from 
first two terms on the right hand side in Eq. \ref{eq:Hint}. Hence the total
transition is $l=0$ to $l=1$ along with a flip of the proton spin. We take the
final state to be $j=3/2$ with $j_z=3/2$. 

We point out that there is another contribution to this transition with the
first two terms on right hand side of Eq. \ref{eq:Hint} contributing at the
first interaction and the last term contributing at the second interaction.
This will lead to a transition $l=0$ to $l=1$ to $l=1$ and 
 will involve an $l=1$ intermediate state. This contribution involves 
 the effective momentum of the initial state wave function $k_i$ and is
 suppressed in comparison to the amplitude described in the previous
 paragraph. Furthermore the fusion process takes place through an $l=1$
 state which is suppressed compared to that for $l=0$. Finally the
 intermediate to final state process involves a magnetic transition which
 is known to be suppressed compared to an electric transition. Hence we 
 expect this amplitude to be small and do not consider it further.

The matrix element at the first vertex is given by,
\begin{equation}
	\bra{n}H_I(t)\ket{i}=\frac{i\ e\ \hbar\ g_p\ k_1}{2m_1\sqrt \Omega}\sqrt{\frac{\hbar}{2\omega_1}}\bra{n}\vec \sigma\cdot(\hat k_1\times\hat \epsilon^*)\ a^\dagger\ e^{-i\ \vec k_1\cdot\vec r+i\omega_1t}\ket{i}
\end{equation}
There is one more term in this proportional to $1/m_2$ which is much smaller
in the limit $m_2>>m_1$
and hence has been neglected \cite{Jain2020}. Furthermore the exponent $\vec k_1\cdot \vec r$ has a multiplicative  
factor $m_2/(m_1+m_2)$, which has been approximated as unity in the limit 
$m_2>>m_1$. 
Let us express the unit vector $\hat k_1$ as,
\begin{equation}
	\hat k_1 = \cos\theta_1\hat z + \sin\theta_1\left[\cos\phi_1\hat x
	+\sin\phi_1 \hat y\right]
	\label{eq:k1}
\end{equation}
We take the two photon polarization vectors to be
\begin{eqnarray}
	\hat\epsilon_1 &=& -\sin\theta_1\hat z + 
\cos\theta_1\left[\cos\phi_1\hat x +\sin\phi_1 \hat y\right] \nonumber\\
	\hat\epsilon_2 &=& -\sin\phi_1\hat x +\cos\phi_1 \hat y 
	\label{eq:eps1}
\end{eqnarray}
We consider the contribution only from $\hat\epsilon_1$. This is sufficient
for our purpose since the contribution from $\hat\epsilon_2$ will add 
incoherently and can only change the result by a factor of order unity.
In any case the other polarization does not contribute to the particular
transition being considered. 
Using $\hat k_1\times \hat\epsilon_1 = \hat\epsilon_2$, we obtain, 
\begin{equation}
\vec s\cdot (\hat k_1\times \hat \epsilon_1)  = \cos\phi_1 s_y - \sin\phi_1 s_x 
\end{equation}
where $\vec s = \vec \sigma/2$. 

The matrix element in the right hand side can now be written as, 
\begin{equation}
	\bra{n}\vec \sigma\cdot(\hat k_1\times\hat \epsilon^*)\ a^\dagger\ e^{-i\ \vec k_1\cdot\vec r+i\omega_1t}\ket{i} = \bra{n}e^{-i\ \vec k_1\cdot\vec r+i\omega_1t}\ket{i}\ (-i\cos\phi_1 - \sin\phi_1) 
\end{equation}
where the spin part has been evaluated. 
In the spatial part of the matrix element, only the $l=0$ term of the photon wave function contribute. Hence using plane wave expansion and doing the angular integral, we obtain,
\begin{equation}
	\bra{n}e^{-i\ \vec k_1\cdot\vec r+i\omega_1t}\ket{i}=  e^{i\omega_1 t}
	\int dr\ U^*_n\frac{\sin( k_1 r)}{ k_1 r}U_i
\end{equation}
In this evaluation we have assumed that the photon with frequency $\omega_1$ 
is emitted at the first vertex. As shown in Fig. \ref{fig:twophotons}, there are
two contributions and either of the photons can be emitted at the
first vertex. However, for the range of parameters considered, 
the second contribution corresponding to photon of frequency $\omega_2$
emitted at the first vertex turns out to be much smaller. This is because
of our choice of $k_1$ and $k_2$ values and is discussed later in section
\ref{sec:results}.

We next consider the matrix element from intermediate to final state. 
The emitted photon has wave vector $\vec k_2$ and frequency $\omega_2$. We 
denote the polar coordinates of the unit vector $\hat k_2$ by $\theta_2$
and $\phi_2$. The polarization vectors are denoted by $\vec \epsilon{\,'}_1$
and $\vec \epsilon{\,'}_2$. These three unit vectors are given by Eqs. \ref{eq:k1}
and \ref{eq:eps1} with $(\theta_1,\phi_1)$ replaced by $(\theta_2,\phi_2)$
We consider the nuclear final state with $j=3/2$ and $j_z=3/2$. The other 
states will add incoherently and including them will only produce a change
of order unity. Hence, we ignore them here to focus on the main result.
Furthermore we take the polarization vector of the photon produced at this
vertex to be $\vec\epsilon{\,'}_2$.
The matrix element evaluates to
\begin{eqnarray}
	\bra{f}H_I(t)\ket{n} &=& -e^{i\omega_2 t}ie \sqrt{\frac{1}{2\Omega \omega_2\hbar}} (E_f-E_n) \bra{f} \vec\epsilon{\,'}_2\cdot\vec r\ket{n}\nonumber\\
	& =&	-e^{i\omega_2 t}ie \sqrt{\frac{1}{12\Omega \omega_2\hbar}}
	(E_f-E_n)(\sin\phi_2+i\cos\phi_2) \int dr U^*_f r U_n
\end{eqnarray}
The reaction rate can be expressed as:
\begin{equation}
{dP\over dt} = {1\over \Delta T} \int dE_1 dE_2 \rho_1\rho_2 
|\langle f|T(t_0,t)|i\rangle|^2
\end{equation}
where $E_1=\hbar \omega_1$, $E_2=\hbar \omega_2$ and 
$\rho_1$ is the photon density of state factor,
given by,
\begin{equation}
    \rho_1 = {\Omega\omega_1^{ 2}\over (2\pi)^3} {d\phi_1 d\cos\theta_1\over \hbar c^3}
    \label{eq:photonDOS}
\end{equation}
along with a corresponding formula for $\rho_2$. 
This leads to the following formula for the reaction rate 
\begin{equation}
    \frac{dP}{dt} = \frac{\alpha^2 g_p^2}{12\pi\hbar^3 c^6m_p^2}\int dE_1\ E_1^3\ E_2\ |I|^2
\end{equation}
\begin{equation}\label{eq:inter}
    I=\sum_nI_1I_2\frac{E_f-E_n}{E_n-E_i+E_1}
\end{equation}
\begin{equation}
    I_1=\int dr\ U_n^*\frac{\sin ( k_1r)}{ k_1 r}U_i
	\label{eq:I1}
\end{equation}
\begin{equation}\label{eq:I2}
    I_2=\int dr'\ U_f^*\ r'\ U_n
\end{equation}
where $ k_1$ is the photon wave number. In Eq. \ref{eq:inter}, the sum over intermediate states runs over all the energy values from zero till infinity. 
We next turn to the calculation of the integrals $I_1$, $I_2$ and the sum
over energies. We point out that it is the sum over energies which led 
to the acute cancellation observed in \cite{Jain2020,Jain2021}.

\subsection{Calculational Details}
In this section we provide some details of the 
compututation of the molecular matrix element, i.e. the integral
$I_1$, given in Eq. \ref{eq:I1}. 
Both the wave functions $U_1$ and $U_n$ correspond to $l=0$. 
For $r<L_b$ the wave function $U_i$ decays very sharply. Hence the dominant
contribution is obtained from $L_c>r>L_b$, although in our calculation
we include contribution from all regions.
The wave functions in the region $L_c>r>L_b$ 
can be expressed as,
\begin{equation}\label{intermediate}
    U_n = \frac{1}{K_nN_n}\ \left[ D_n\ \sin(K_n\ r) + F_n\ \cos(K_n\ r)\right]
\end{equation}
\begin{equation}\label{initial}
    U_i = \frac{1}{K_iN_i}\ \left[ D_i\ \sin(K_i\ r) + F_i\ \cos(K_i\ r)\right]
\end{equation}
Using these equations and simplifying $I_1$ we get a sum of 4 sine terms and 4 cosine terms, given below, 
\begin{eqnarray}
	T_1&=& \sin\left(( k_1+K_n-K_i)r\right)\ [D_n D_i + F_n F_i]
	\nonumber\\
    	T_2&=& \sin\left(( k_1-K_n+K_i)r\right)\ [D_n D_i + F_n F_i]
	\nonumber\\
  	T_3&=&   \sin\left(( k_1+K_n+K_i)r\right)\ [F_n F_i - D_n D_i]
	\nonumber\\
   	T_4&=&  \sin\left(( k_1-K_n-K_i)r\right)\ [F_n F_i - D_n D_i]
	\nonumber\\
  	T_5&=&   -\cos\left(( k_1+K_n-K_i)r\right)\ [D_n F_i - F_n D_i]
	\nonumber\\
    	T_6&=& -\cos\left(( k_1-K_n+K_i)r\right)\ [F_n D_i - D_n F_i]
	\nonumber\\
    	T_7&=& \cos\left(( k_1+K_n+K_i)r\right)\ [-F_n D_i - D_n F_i]
	\nonumber\\
    	T_8&=& \cos\left(( k_1-K_n-K_i)r\right)\ [F_n D_i + D_n F_i]
	\label{eq:Ti}
\end{eqnarray}
each with a pre-factor,
\[\frac{1}{4 N_i K_i}\frac{1}{N_n K_n}\frac{1}{ k_1}\frac{1}{r}\]
For small $K_i$ the dominant contribution comes from 
the kinematic region $K_n\approx k_1$ due to 
terms which involve $K_n- k_1$.
The sum of these two wave numbers is very large and leads to neglible values
of $I_1$. 

The integrals are facilitated by the following formulas,
\begin{equation}
	\int_{L_b}^{L_c} \frac{\sin(\mathcal K\ r)}{r} dr = 
	\left[{\rm Si}(|\mathcal K|\ L_c)  -
	{\rm Si}(|\mathcal K|\ L_b)\right]\ {\rm sgn}(\mathcal K)
\end{equation}
where 
\begin{equation}
	{\rm Si}(x)  = \int_0^x dt\ {\sin t\over t}
\end{equation}
Similarly the cosine integral gives,
\begin{equation}
	\int_{L_b}^{L_c} \frac{\cos(\mathcal K\ r)}{r} dr =
	{\rm Ci}(|\mathcal K|\ L_c)-{\rm Ci}(|\mathcal K|\ L_b)
\end{equation}
where ${\rm Ci}(x)$ can be expressed as,
\begin{equation}
	{\rm Ci}(x) = \gamma+\ln x - \int_0^x{1-\cos t\over t}dt
	\label{eq:Ci}
\end{equation}
We similarly compute the integral in the region $r<L_b$. 

We next consider the integral $I_2$ which appears in the nuclear 
matrix element. 
This integral gets dominant contribution from very small values of $r$. 
It shows a very mild dependence on
intermediate state energy eigenvalue $E_n$ for
$E_n<V_1$, 
except for the factor $N_n$ in the denominator. This essentially 
corresponds to a barrier penetration factor and shows an increase
with $E_n$.
For $E_n>> V_1$, $I_2$ starts to oscillate with $E_n$.

\section{Results}
\label{sec:results}

We first consider the case $L_c=\infty$. In this case the potential $V_2$ plays
no role  and we obtain a continuous energy spectrum. We take the initial energy
eigenvalue $E_i$ to be equal to 0.1 eV. As found earlier \cite{Jain2020,Jain2021} the amplitude 
becomes large in the regime when $K_n\approx k_1$. However adding over all
intermediate eigenstates leads to a very small result. The precise value is
difficult to compute numerically but we find a cancellation up to six 
significant digits. This suggests that the rate in free space is 
very small. This result is in agreement with what was found in 
\cite{Jain2020,Jain2021}. We also test the possibility that rate may be 
enhanced if the photon wave number $k_1$ is set equal to the wave number
$K_n$ of
 an intermediate state which corresponds to a nuclear resonance. We find that even
in this case the rate turns to be very small due to a delicate cancellation.

\begin{figure}[H]
\centering
\includegraphics[ clip,scale=0.7]{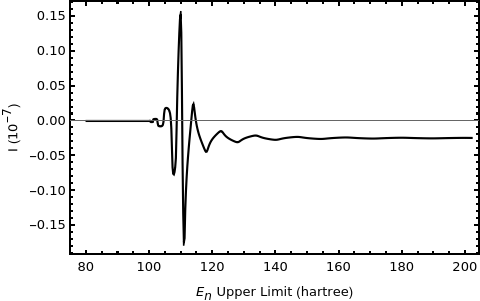}
	\caption{\label{fig:discrete_amp} The amplitude $I$ (Eq. \ref{eq:inter}) 
	as a function of the upper limit on the energy $E_n$ of the 
	intermediate state. Here the photon wave number has been set equal 
	to 635 atomic units and $L_c=10$.
}
	\label{fig:amp_discrete}
\end{figure}

\begin{figure}[H]
\centering
\includegraphics[ clip,scale=0.6]{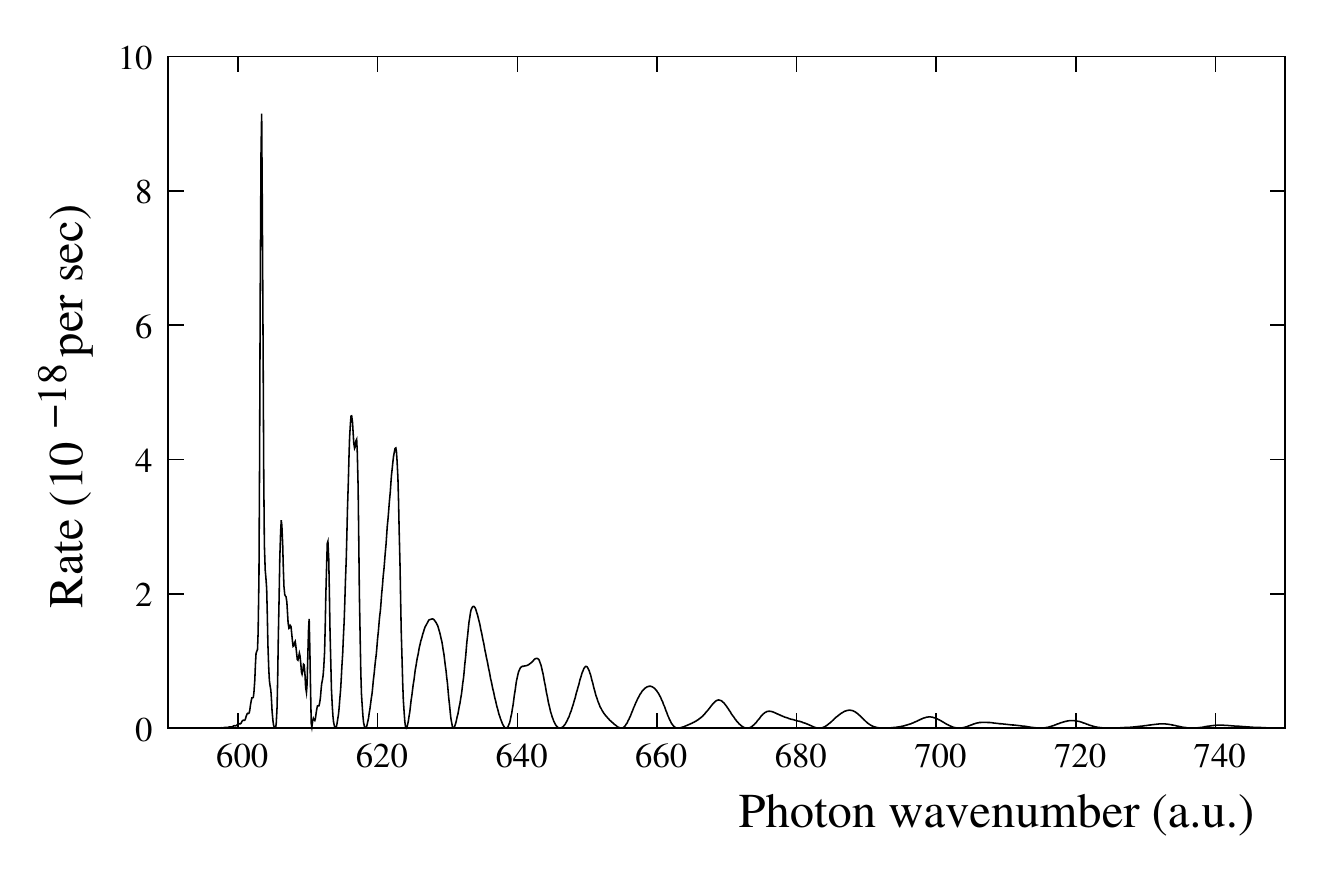}
\caption{\label{fig:rate1} The reaction rate as a function of the wave 
of the photon emitted at the first vertex for $L_c=10$. The rate shows very rapid 
	fluctuations which are apparent in the figure.}
\end{figure}

We next compute the rate for a finite value of $L_c$ which would be
applicable in a medium. We first take a relatively small value of $L_c=10$
in atomic units. This will lead to a relatively large separation in energy 
among different states. 
We may compare this value with the level spacing in the case of 
 Kronig-Penney model which is a one dimensional periodic potential. 
 We find that the level spacing in this case is determined by the width
 of the potential wells 
  \cite{merzbacher1998quantum}. This in our case would represent the
  distance between the lattice sites in the medium and hence would be of
  order a few atomic units.
   Hence the choice $L_c=10$ is not 
  unphysical.  
For this value we find an energy eigenvalue at $E_i=0.0903$
eV which we take to be the initial state energy. We take the photon wavenumber
$k_1$ close to 600 atomic units.
In this case the amplitude is quite large. In Fig. \ref{fig:amp_discrete} 
we show the amplitude $I$, defined in Eq. \ref{eq:inter} as a function
of the upper limit on the intermediate state energy $E_n$ for photon
wave number $k_1=635$ atomic units. We see that
the amplitude settles to a finite value as $E_n\rightarrow \infty$. 
The dependence of rate on the photon 
wavenumber is shown in Fig. \ref{fig:rate1}.
We find a rate of the order of $10^{-18}$ 
per second for the photon wave number close to 600 atomic units. 
The rate in this case
is much larger than that found for the first order transition.
If we 
set the chosen photon
wave number equal to intermediate state wave number $K_n$, the corresponding
intermediate energy comes out to be close to 100 atomic units, which is 
equal to the barrier height $V_1$. 
From Fig. \ref{fig:rate1} it is clear that this choice of wave number
leads to maximum rate. 
We also check the contribution due to the second term in Fig. \ref{fig:twophotons} with $k_1$ interchanged
with $k_2$. For $k_1=600$, $k_2$ is close to 3000. For these two interchanged
the amplitude is found to be four orders of magnitude smaller and hence
negligible. We can, therefore, ignore the second amplitude for this choice of 
parameters. 

\begin{figure}[H]
\centering
\includegraphics[ clip,scale=0.6]{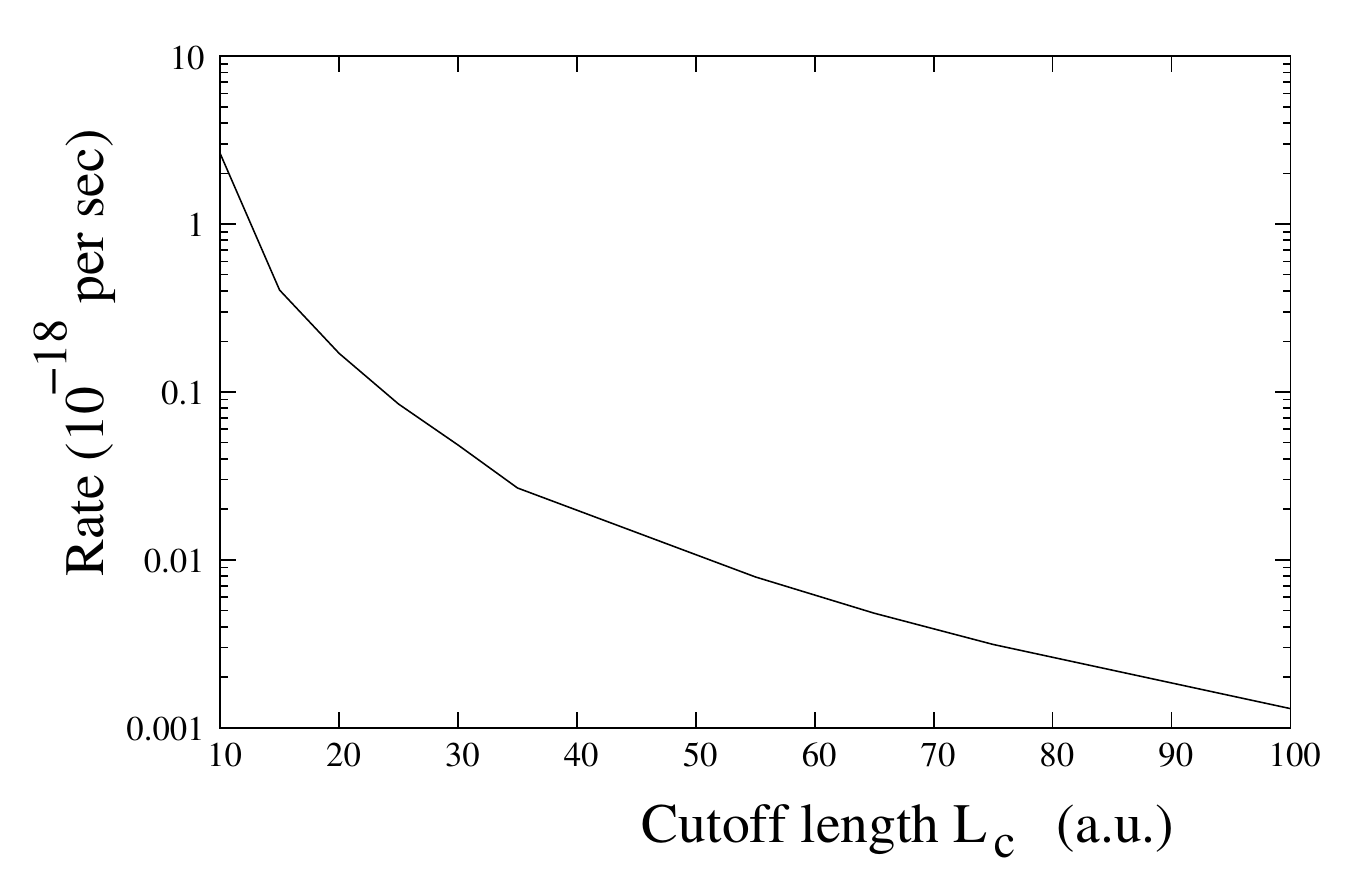}
\caption{\label{fig:rate2} The reaction rate as a function of the 
cutoff length scale $L_c$ for photon wave number $k_1=635$ atomic
	units. We find that the rate drops roughly as
$1/L_c^3$. 
	}
\end{figure}

As we increase the value of $L_c$ we find that the dominant change occurs
due to the normalization factors in the wave functions. Both the $U_i$ and 
$U_n$ wave functions contain a factor $\sqrt{L_c}$ in the denominator. 
Hence the rate shows an approximate decrease as $1/L_c^3$. 
In Fig. \ref{fig:rate2} we show the dependence of rate on $L_c$ for
$k_1=635$ atomic units for $L_c$ up to 100. We have explicitly checked
the dependence on $L_c$ up to $L_c=1000$ and the trend shown in
Fig. \ref{fig:rate2} continues up to this value. At some sufficiently 
large value of $L_c$ we expect a much sharper decrease, due to a delicate
cancellation that was seen in the continuum case. However the numerical work
to obtain this value of $L_c$ becomes prohibitively time intensive 
and we do not pursue this in the current paper.

\section{Conclusions}
We conclude that nuclear fusion reactions may be possible at low energies inside
a medium 
through the mechanism of second order perturbation theory 
\cite{Jain2020,Jain2021,PhysRevC.99.054620}. 
We find that the rate in free space turns out to be very small, in agreement
with earlier results \cite{Jain2020,Jain2021}. The amplitude in this case
does become relatively large for intermediate state wave number values $K_n$
close to the photon
wave number $k_1$. However as we sum over all intermediate wave numbers the 
amplitude becomes very small due to a delicate cancellation. In a medium,
however, we expect that the energy eigenvalues would not be continuous. 
This is seen, for example, in a crystalline lattice which leads to a
band structure with energy levels being separated by forbidden bands
\cite{Ashcroft76}. 
In a disordered system we expect to see localized states, analogous to 
Anderson localization \cite{doi:10.1142/7663,RevModPhys.57.287}.
In either case we expect discontinuous energy eigenvalues. In the present paper
we use a simple model by imposing a hard wall cutoff beyond a certain
length scale. The essential feature of this model is that it
leads to discretization of energy levels. This model leads to substantial 
rates for fusion reaction to take place at second order in perturbation
theory. We obtain dominant contribution from intermediate states whose
wave number $K_n$ in the region $L_b<r<L_c$ is close to the photon wave number
$k_1$. The contribution is maximal if the intermediate states have
energy close to the height of the potential barrier. 

Our results suggest that, in favorable conditions,
nuclear fusion reactions can take place at low
energies at observable rates. However, so far we have only presented a 
toy model to test this phenomenon. Considerable more effort is needed 
to make contact with observations. This will require use of realistic
models of the solid structure. Furthermore, photon emission may not be
the dominant process and it would be interesting to consider other mechanisms.

\bibliographystyle{unsrt}
\bibliography{nuclear}
\end{document}